\documentclass[a4paper,12pt]{article}

\newcommand{\sect}[1]{\setcounter{equation}{0}\section{#1}}

\begin{document}
\topmargin 0pt
\oddsidemargin 0mm

\renewcommand{\thefootnote}{\fnsymbol{footnote}}
\begin{titlepage}
\begin{flushright}
OU-HET 357 \\
hep-th/0008119
\end{flushright}

\vspace{5mm}
\begin{center}
{\Large \bf Lorentz Transformation and Light-Like Noncommutative SYM}
\vspace{10mm}

{\large
Rong-Gen Cai\footnote{e-mail address: cai@het.phys.sci.osaka-u.ac.jp} and
Nobuyoshi Ohta\footnote{e-mail address: ohta@phys.sci.osaka-u.ac.jp}} \\
\vspace{8mm}
{\em Department of Physics, Osaka University,
Toyonaka, Osaka 560-0043, Japan}

\end{center}
\vspace{5mm}
\centerline{{\bf{Abstract}}}
\vspace{5mm}

We show that combining the spatial noncommutative SYM limit and Lorentz
transformation, one can obtain a well-behaved light-like noncommutative
SYM limit. The light-like noncommutative SYM is unitary. When the boost
velocity is finite, the resulting theory with space-time noncommutativity is
unitary as well. The light-like noncommutative SYM limit can also be approached
by combining the noncommutative open string theory limit and Lorentz
transformation. Along this line, we obtain the supergravity dual for the
light-like noncommutative SYM, which is the same as the one acquired using
a different method. As a comparison, the supergravity duals for the ordinary
SYM, spatial noncommutative SYM and the noncommutative open string theories
are given as well, in an infinitely-boosted frame with finite momentum density,
which are the decoupling limits of bound states (D$p$, W), (D($p-2$), W, D$p$),
and (F$1$, W, D$p$), respectively.

\end{titlepage}

\newpage
\renewcommand{\thefootnote}{\arabic{footnote}}
\setcounter{footnote}{0}
\setcounter{page}{2}

\sect{Introduction}
Over the past years there have been a lot of activities on the low energy
decoupling limits of string theory involving D$p$-branes. Initially,
Maldacena \cite{Mald1}
found that in a certain low energy limit (SYM limit) the closed strings and
massive modes of open strings ending on the D$p$-branes decouple, one is
left with a field theory (super Yang-Mills theory) and such a low energy
field theory with strong 't Hooft coupling has a weak coupling dual
description: supergravity. In particular, string/M theory on
an anti-de Sitter space (AdS) with a compact space is conjectured to be
dual to a certain large $N$ conformal field theory (CFT) which lives on the
boundary of AdS \cite{Mald1,Gubs,Witt1,Itzh}.

Later it was found that when the D$p$-brane is put in a background with a
non-zero NS $B$ field, the worldvolume coordinates become noncommutative
\cite{Dougl,Chu,Arda}. For a D$p$-brane with only spatial component $B$ field,
one can also find a limit (NCSYM limit), in which the closed strings and
massive modes of open strings are decoupled, and ends up with a SYM in a
space-space noncommutative spacetime, which is called noncommutative SYM
(NCSYM) \cite{Seib1}. There also exists a supergravity dual description for
such spatial NCSYM's \cite{Mald2}-\cite{Youm},
from which one can investigate a lot of properties of NCSYM's. In contrast
to the case of spatial component NS $B$ field, if the D$p$-brane is put in
a non-zero NS $B$ field with only a time-like component, it was found that
one cannot define a low energy field theory limit \cite{Seib2,Gopa1,Barb2}.
Instead one can find a decoupling limit (NCOS limit), in which the closed
strings decouple and one is left with a open string theory in a space-time
noncommutative spacetime (NCOS). Since then, there have been a lot of papers
appearing on the net, including \cite{Seib3}-\cite{LRS2}, in which
supergravity duals, the relation to the NCSYM and other related topics
have been discussed.

It is argued that field theories with space-time noncommutativity may
exhibit acausal behavior \cite{Seib3,Alva}, and quantum field theories on
such a spacetime are not unitary \cite{Gomi}. However, most recently,
Aharony, Gomis and Mehen \cite{AGM} have shown that quantum field theories
with light-like noncommutativity are unitary and they can be achieved through
a certain low energy limit of string theory. Subsequently Alishahiha, Oz, and
Russo (AOR)\cite{AOR} have constructed the dual supergravity description of
field theories with light-like noncommutativity.

In this paper we will further consider quantum field theories with light-like
noncommutativity and their dual supergravity description. In order to get
supergravity duals for the SYM with light-like noncommutativity,
AOR first constructed D$p$-brane solutions with
a non-zero light-like background NS $B$ field and then took a usual SYM
limit for that goal. To clearly see the relation of light-like
NCSYM to the spatial commutative SYM and to the NCOS, we get the
light-like NCSYM limit by combining the Lorentz transformation and the
NCSYM decoupling limit or the NCOS decoupling limit. In sect.~2
we consider the Lorentz transformation, the NCSYM limit and the NCOS
limit, and show that a well-behaved light-like NCSYM limit is obtained.
The case for the finite boost velocity is also discussed. Along this
line, in sect.~3 we re-derive the supergravity dual for the light-like
NCSYM. In sect.~4 we consider the Lorentz transformation, along other
directions, of the supergravity duals for the ordinary SYM (OSYM),
the spatial NCSYM and the NCOS, thereby we obtain the supergravity duals
for the OSYM, the spatial NCSYM and the NCOS in an
infinitely-boosted frame with finite momentum density. The conclusion is
included in sect. 5.


\sect{Lorentz transformation and decoupling limits }

To discuss a low energy decoupling limit of string theory involving
D$p$-branes, a good starting point is the Seiberg-Witten
relation connecting the open and closed string moduli \cite{Seib1}:
\begin{eqnarray}
&& G_{ij}=g_{ij}-(2\pi\alpha')^2 (Bg^{-1}B)_{ij}, \nonumber\\
&& \Theta^{ij}=2\pi \alpha' \left(\frac{1}{g+2\pi\alpha' B}
   \right)^{ij}_A, \nonumber\\
&& G^{ij}=\left(\frac{1}{g + 2\pi \alpha' B}\right)_S^{ij},\nonumber\\
&& G_s =g_s\left (\frac{\det G_{ij}}
   {\det(g_{ij} +2\pi\alpha' B_{ij})}\right)^{1/2},
\end{eqnarray}
where $(\;)_A$ and $(\;)_S$ denote the antisymmetric and symmetric parts,
respectively. The open string moduli occur in the disk correlators on the
open string worksheet boundaries
\begin{equation}
<X^i(\tau)X^j(0)> = -\alpha' G^{ij} \ln(\tau)^2 +\frac{i}{2}\Theta ^{ij}
 \epsilon (\tau).
\end{equation}

For simplicity of notation, we restrict ourselves here to the case of
D3-branes. It is straightforward to extend to other dimensions. Suppose
we have the following closed string metric:
\begin{equation}
\label{coor}
ds^2 =-dx_0^2 +dx^2_1 +g (dx_2^2 +dx_3^2).
\end{equation}
Here we have written down only the worldvolume coordinate part.

\subsection{From the spatial NCSYM to the light-like NCSYM}

When the NS $B$ field has only nonvanishing spatial component, one has a
well-defined decoupling limit, and the resulting theory is the spatial
NCSYM. For example, suppose the $B$ field has only constant
component along $x_2-x_3$ directions,
\begin{equation}
B_{ij}=B dx_2\wedge dx_3.
\end{equation}
Rescaling the closed string metric $g$ and the closed string coupling
constant $g_s$ by
\begin{equation}
\label{ncsym}
g=(2\pi \alpha' B)^2,\ \ \
g_s=2\pi \alpha'B G_s,
\end{equation}
and keeping $B$ as a finite constant, in the decoupling limit $\alpha' \to 0$,
we have
\begin{equation}
G^{ij}=\eta^{ij}, \ \ \ \Theta^{ij}=B^{-1}(-\delta^i_2\delta^j_3
 +\delta^i_3\delta^j_2).
\end{equation}
Thus we obtain a 3+1 dimensional SYM in a spacetime with space-space
noncommutativity $\Theta^{23}=-B^{-1}$. This field theory is shown to be
unitary \cite{Gomi}.

Now we perform a Lorentz transformation for the coordinate (\ref{coor})
\begin{eqnarray}
&& x_0=\cosh \gamma x_0'-\sinh\gamma x_2', \nonumber\\
\label{lorentz}
&& x_2=-\sinh\gamma x_0'+\cosh\gamma x_2'.
\end{eqnarray}
Using the Seiberg-Witten relation, we have
\begin{eqnarray}
&& G^{ij} = \frac{1}{g^2 +4\pi^2 \alpha'^2 B^2} \times \nonumber \\
&& \left (
\begin{array}{cccc}
{\scriptstyle -(g^2+4\pi^2 \alpha'^2 B^2)\cosh^2\gamma +g \sinh^2\gamma}
& 0 & {\scriptstyle (g-g^2-4\pi^2 \alpha'^2 B^2) \cosh\gamma\sinh\gamma}
& 0 \\
0 & {\scriptstyle g^2 +4\pi^2 \alpha'^2 B^2} & 0 &0 \\
{\scriptstyle (g-g^2-4\pi^2 \alpha'^2 B^2) \cosh\gamma\sinh\gamma} & 0
& {\scriptstyle g\cosh^2\gamma -(g^2 +4\pi^2 \alpha'^2 B^2) \sinh^2\gamma}
 &0 \\
0 & 0& 0& g
\end{array}
\right )
\end{eqnarray}
and the noncommutativity matrix
\begin{equation}
\Theta^{ij} =\frac{2\pi\alpha'}{g^2 +4\pi^2 \alpha'^2 B^2}
\left (
\begin{array}{cccc}
0 & 0& 0& -2\pi \alpha' B \sinh\gamma \\
0& 0& 0& 0 \\
0& 0& 0& -2\pi\alpha' B \cosh\gamma\\
2\pi \alpha' B\sinh\gamma &0& 2\pi\alpha' B \cosh\gamma &0
\end{array}
\right).
\end{equation}
Taking the same NCSYM limit (\ref{ncsym}), we reach
\begin{equation}
G^{ij}=\eta^{ij},
\end{equation}
and
\begin{equation}
\label{ncm1}
\Theta^{ij}= \left (
\begin{array}{cccc}
0 & 0& 0& -B^{-1}\sinh\gamma \\
0 & 0& 0& 0 \\
0 & 0& 0& -B^{-1}\cosh\gamma \\
B^{-1}\sinh\gamma &0& B^{-1}\cosh\gamma &0
\end{array}
\right).
\end{equation}
Obviously, this is a low-energy field theory limit with space-time
noncommutativity when $\gamma$ is finite. We will argue that field theory thus
defined is unitary even though the space-time coordinates are noncommutative.

On the other
hand, when the boost velocity approaches the speed of light, that is,
$\gamma \to \infty$, the Lorentz transformation (\ref{lorentz}) is
singular and some components of the noncommutativity matrix (\ref{ncm1})
are divergent. But we notice that rescaling the constant $B$ as
\begin{equation}
e^{\gamma}/B = b,
\end{equation}
where $b$ is a finite constant, yields a well-behaved
noncommutativity matrix with space-time component,
\begin{equation}
\Theta^{ij}= \left (
\begin{array}{cccc}
0 & 0& 0& -b \\
0 & 0& 0& 0 \\
0 & 0& 0& -b \\
b &0& b &0
\end{array}
\right).
\end{equation}
Actually, this limit is just the low energy field theory limit with
light-like noncommutativity discussed in \cite{AGM}. There it is
shown that this theory is unitary despite the nonlocality in the
time coordinate. This result is reasonable since this low-energy field
theory is an infinitely-boosted limit of a unitary theory.

\subsection{From the NCOS to the light-like NCSYM}

Now it is clear that when the NS $B$ field has a nonvanishing time-like
component, one can define a limit, in which the closed strings are decoupled
and one ends up with a noncommutative open string theory \cite{Seib2,Gopa1}.
 For example, suppose the $B$ field
has, in the coordinate (\ref{coor}), the time-like component as
\begin{equation}
B_{ij}=E dx_0 \wedge dx_1.
\end{equation}
Rescaling the electric field $E$, the closed string metric $g$ and the
closed string coupling constant $g_s$ as
\begin{equation}
\label{ncos}
2\pi \alpha' E =1-\frac{1}{2}\frac{\alpha'}{\alpha'_{eff}},\ \ \
 g=\frac{\alpha'}{\alpha'_{eff}}, \ \ \ g_s=G_s\sqrt{\frac{\alpha'_{eff}}
{\alpha'}},
\end{equation}
where $\alpha'_{eff}$ is a finite constant, in the decoupling limit
$\alpha' \to 0$, one has
\begin{equation}
\alpha' G^{ij}=\alpha'_{eff}\eta^{ij}, \ \ \ \Theta^{ij}=2\pi \alpha'_{eff}
 (\delta^i_0\delta^j_1-\delta_1\delta^j_0).
\end{equation}
This is a NCOS limit with the open string tension $1/4\pi\alpha'_{eff}$, the
coupling constant $G_s$ and the space-time noncommutativity $\Theta ^{01}
= 2\pi \alpha'_{eff}$.

Now we make also the same Lorentz transformation (\ref{lorentz}). In this case,
we find
\begin{equation}
G^{ij} = \left (
\begin{array}{cccc}
-\frac{g\cosh^2\gamma -(1-e^2)\sinh^2\gamma}{g(1-e^2)} & 0&
-\frac{(g-1+e^2)\sinh\gamma\cosh\gamma}{g(1-e^2)} &0 \\
0 &\frac{1}{1-e^2} &0 &0 \\
-\frac{(g-1+e^2)\sinh\gamma \cosh\gamma}{g(1-e^2)} &0& \frac{(1-e^2)
\cosh^2\gamma-g\sinh\gamma}{g(1-e^2)}&0 \\
0 & 0& 0& \frac{1}{g}
\end{array} \right ),
\end{equation}
where $e=2\pi \alpha' E$, and the noncommutativity matrix
\begin{equation}
\Theta^{ij}=\frac{2\pi \alpha'e}{1-e^2} \left (
\begin{array}{cccc}
0 & \cosh\gamma &0& 0\\
-\cosh\gamma &0 &-\sinh\gamma &0 \\
0& \sinh\gamma &0 &0 \\
0 & 0& 0& 0
\end{array} \right ).
\end{equation}
Taking the same NCOS decoupling limit (\ref{ncos}), we obtain
\begin{equation}
\label{metric}
\alpha' G^{ij} =\alpha'_{eff}\eta ^{ij},
\end{equation}
and
\begin{equation}
\label{theta}
\Theta^{ij}= 2\pi \alpha'_{eff} \left (
\begin{array}{cccc}
0 & \cosh\gamma &0& 0\\
-\cosh\gamma &0 &-\sinh\gamma &0 \\
0& \sinh\gamma &0 &0 \\
0 & 0& 0& 0
\end{array} \right ).
\end{equation}
Obviously, when the boost velocity is finite, this is a well-behaved NCOS
limit. It means that the NCOS is still a NCOS with both the space-time and
space-space noncommutativities after the Lorentz
transformation \cite{Chen}. Of course,
the space-space noncommutativity can be removed using a coordinate
transformation. However, when the boost velocity
approaches the speed of light, $\gamma \to \infty$, from the
noncommutativity matrix (\ref{theta}) one can see that in order to have a
finite noncommutativity value, one has to rescale the constant $\alpha'_{eff}$
 as
\begin{equation}
\label{rescaling}
\alpha'_{eff}e^{\gamma} =b',
\end{equation}
where $b'$ is another finite constant. In this case, we see from (\ref{metric})
that $ \alpha' G^{ij}=0$. It means that when the boost velocity approaches
the speed of light, from the NCOS limit we can define a well-behaved low energy
field theory limit. Actually, this limit is also the field theory
limit with light-like noncommutativity. Furthermore, equations (\ref{metric})
and (\ref{rescaling}) implies that in order to have a well-defined open string
metric, one has to further rescale all coordinates as
\begin{equation}
x_i \sim e^{\gamma/2}x_i.
\end{equation}
This will be used later when we discuss the supergravity dual.

\subsection{Unitarity}

In \cite{Gomi} the unitarity of quantum fields has been studied in
a noncommutative spacetime. It has been found that space-like
noncommutative quantum field theories are always unitary while time-like
noncommutative quantum theories are not unitary. In \cite{AGM} it was
shown that quantum theories with light-like noncommutativity are unitary
as well.

The unitarity of quantum field theories requires that the inner product
$p\circ p$ be never negative \cite{Gomi,AGM}, where $p$ is some external
momentum and
\begin{equation}
p \circ p \equiv -p_{\mu} \Theta^{\mu \rho} G_{\rho\sigma}\Theta^{\sigma\nu}
 p_{\nu} \equiv p_{\mu} g^{\mu\nu}_{\theta}p_{\nu} \ge 0,
\end{equation}
where $\Theta^{\mu\nu}$ is the noncommutativity matrix and $G_{\rho\sigma}$
is the background metric of quantum fields. Let us first consider the spatial
NCSYM in 3+1 dimensions discussed above. In that case,
we have
\begin{equation}
p \circ p = B^{-2}(p^2_2 +p_3^2),
\end{equation}
which clearly indicates that the spatial NCSYM satisfies the
requirement of unitarity. Therefore this theory is unitary \cite{Gomi}.
For the finite Lorentz boost, we find
\begin{equation}
p\circ p = B^{-2}(p_0 \sinh \gamma +p_2 \cosh\gamma)^2 + B^{-2}p_3^2.
\end{equation}
Again, the unitarity constraint is satisfied. Therefore we can conclude that the
SYM with time-space noncommutativity, coming from the spatial
NCSYM after Lorentz transformation, is unitary as well.
This seems reasonable because the physics should be invariant after a
regular coordinate transformation. Actually, the two theories are equivalent.
The space-time noncommutativity in the former can be removed using the Lorentz
transformation. Therefore we cannot say in general that quantum field theories
with space-time noncommutativity are not unitary. Of course, the description
in terms of the spatial NCSYM is simpler.

When the boost velocity approaches the speed of light, we have
\begin{equation}
p\circ p = b^2(p_0 +p_2)^2,
\end{equation}
which always satisfies $p\circ p \ge 0$. Thus the SYM with light-like
noncommutativity is unitary. Similarly, the SYM with light-like
noncommutativity coming from the NCOS limit can be proved to be unitary.

\sect{Supergravity duals }

In the previous section we have shown that the low energy SYM with light-like
noncommutativity can be obtained through combining the NCSYM limit and Lorentz
transformation, or the NCOS limit and Lorentz transformation. In this section
we further give evidence to support that point of view. We ``derive'' the
supergravity duals of light-like NCSYM's through the
supergravity duals of spatial NCSYM's or of NCOS's.

Let us first consider the D3-brane case. The supergravity dual of spatial
NCSYM in 3+1 dimensions has been given in \cite{Mald2,Has}, which
is
\begin{eqnarray}
&& ds^2 =\alpha' \left [\frac{u^2}{R^2}\left (-dx_0^2 +dx_1^2 +\tilde{h}
   (dx_2^2 +dx_3^2)\right)
 +\frac{R^2}{u^2}\left( du^2 +u^2 d\Omega_5^2\right)\right], \nonumber \\
&& e^{2\phi}=\hat{g}_s^2 \tilde{h}, \ \ \ B_{23}=\frac{\alpha'}{\tilde{b}}
  \frac{(au)^4}{1+(au)^4},
\end{eqnarray}
where $R^4 =4\pi \hat{g}_s N$, $a^4 =\tilde{b}^2/R^4$, $\tilde{h}^{-1}
=1+(au)^4$, and $N$ denotes the number of D3-branes. The spatial NCSYM
has the coupling constant $g_{YM}^2=2 \pi\hat{g}_s$.
 Performing the coordinate transformation (\ref{lorentz}) for coordinates
$x_0$ and $x_3$, we have
\begin{eqnarray}
ds^2 &=& \alpha' \left [\frac{u^2}{R^2}\left (-dx_0^2 +dx_1^2 + \tilde{h}
 dx_2^2 +dx_3^2 +(\tilde{h}-1) (\cosh\gamma dx_3^2 -\sinh\gamma dx_0)^2
  \right) \right. \nonumber \\
 && ~~~~~~\left. +\frac{R^2}{u^2}\left (du^2 +u^2 d\Omega_5^2\right)\right].
\end{eqnarray}
This supergravity solution is supposed to be the supergravity dual of the
spatial NCSYM after a finite Lorentz boost considered in the
previous section. In this coordinate, space-time coordinates are noncommutative.
When $\gamma \to \infty$, rescaling the constant $\tilde{b}$ as
\begin{equation}
\label{b}
\tilde{b} e^{\gamma} = b,
\end{equation}
we have
\begin{equation}
\label{lightsym}
ds^2 =\alpha' \left[ \frac{u^2}{R^2}\left (dx_+ dx_- +dx_1^2 +dx_2^2
  -\frac{b^2 u^4}{R^4}dx_-^2\right) +\frac{R^2}{u^2}
 \left (du^2 +u^2 d\Omega_5^2\right)\right],
\end{equation}
where $x_{\pm}= x_3 \pm x_0$. The dilaton and $B$ field become
\begin{equation}
e^{2\phi}=\hat{g}_s^2, \ \ \ B_{2-}=\alpha' \frac{b u^4}{R^4}.
\end{equation}
We see that this supergravity solution is just the supergravity dual
of light-like NCSYM found in \cite{AOR}. The light-like
NCSYM has the coupling constant $g^2_{\rm YM}=2\pi \hat{g}_s$,
the same as the one for the spatial NCSYM.

It is rather easy to extend this to other dimensions. The supergravity dual
of the spatial NCSYM in $p+1$ dimensions is~\cite{CO}
\begin{eqnarray}
&& ds^2=\alpha'\left [\left(\frac{u}{R}\right)^{(7-p)/2}\left (-
  dx_0^2 +dx_1^2 +\cdots +dx_{p-2}^2
  +\tilde{h}(dx_{p-1}^2 +dx_p^2)\right)
   \right. \nonumber \\
&&~~~~~~~~\left. +\left(\frac{R}{u}\right)^{(7-p)/2}
    \left(du^2
 +u^2 d\Omega^2_{8-p}\right)\right], \nonumber\\
\label{ncsymdual}
&& e^{2\phi} = \hat{g}_s^2 \tilde{h}\left(\frac{R}{u}
 \right)^{(7-p)(3-p)/2}, \ \ \ B_{p-1,p}
 = \frac{\alpha'}{\tilde{b}}\frac{(au)^{7-p}}{1+(au)^{7-p}},
\end{eqnarray}
where the corresponding RR fields are not exposed explicitly,
\begin{equation}
 \tilde{h} = \frac{1} {1+(au)^{7-p}}, \ \ \ a^{7-p} = \tilde{b}^2/R^{7-p},
\ \ \ R^{7-p} = \frac{1}{2}(2\pi)^{6-p}\pi^{-(7-p)/2}\Gamma[(7-p)/2]
 \hat{g}_s N.
\end{equation}
Making a similar Lorentz transformation as (\ref{lorentz}) for the
coordinates $x_0$ and $x_{p}$, taking the limit $\gamma \to \infty$
 and rescaling $\tilde{b}$ as (\ref{b}), we obtain
\begin{eqnarray}
&& ds^2=\alpha'\left [\left(\frac{u}{R}\right)^{(7-p)/2}\left (
  dx_+dx_- +dx_1^2 +\cdots +dx_{p-1}^2
 -\frac{b^2 u^{7-p}}{R^{7-p}}dx_-^2 \right)
\right. \nonumber \\
&&~~~~~~~~\left. +\left(\frac{R}{u}\right)^{(7-p)/2}
    \left(du^2
 +u^2 d\Omega^2_{8-p}\right)\right], \nonumber\\
&& e^{2\phi} = \hat{g}_s^2 \left(\frac{R}{u}
 \right)^{(7-p)(3-p)/2}, \ \ \ B_{(p-1)-}
 =\alpha'\frac{b u^{7-p}}{R^{7-p}}.
\end{eqnarray}
Here $x_{\pm}= x_p \pm x_0$. This solution
gives the supergravity dual description of light-like NCSYM in $p+1$
dimensions with coupling constant $g^2_{\rm YM}=(2\pi)^{p-2}\hat{g}_s$.

On the other hand, from the previous section we see that the light-like
NCSYM limit can also be achieved from the NCOS limit combining
the Lorentz transformation. Now we also get the supergravity duals of
the light-like NCSYM's along this line.

The supergravity dual for the 3+1 dimensional NCOS has been given
in \cite{Gopa1}. It can be described as
\begin{eqnarray}
&& ds^2 = \alpha' F^{1/2}\left [ \frac{u^4}{R^4}\left (-dx_0^2 +dx_1^2 \right )
  +F^{-1}\left (dx_2^2 +dx_3^2\right ) +du^2 +u^2 d\Omega^2_5 \right ],
 \nonumber \\
 && F=1+\frac{R^4}{u^4},\ \ \ e^{2\phi}= \hat{g}_s^2 F \frac{u^4}{R^4}, \ \ \
 B_{01}=\alpha' \frac{u^4}{R^4},
\end{eqnarray}
where $R$ is the same as the corresponding one above.
Performing a Lorentz transformation (\ref{lorentz}) for the coordinates
$x_0$ and $x_3$, one has
\begin{eqnarray}
 ds^2 &=& \alpha' F^{1/2}\left [ \frac{u^4}{R^4}\left (-dx_0^2 +dx_1^2 +
 dx_3^2\right) + F^{-1}dx_2^2 \right. \nonumber \\
 && ~~~~~~~ \left. -\frac{u^8}{R^4(R^4+u^4)}
 \left(\cosh \gamma dx_3 -\sinh\gamma dx_0 \right )^2
 + du^2 + u^2 d\Omega_5^2 \right ].
\end{eqnarray}
Taking the limit $\gamma \to \infty$ and rescaling coordinates as
\begin{equation}
\label{u}
u \to u e^{-\gamma /2}, \ \ \ \ x_i \to x_i e^{\gamma/2},
\end{equation}
we reach
\begin{equation}
ds^2 =\alpha' \left [\frac{u^2}{R^2}\left (dx_+dx_- + dx_1^2 +dx_2^2
 -\frac{u^4}{R^4}dx_-^2\right ) +
 \frac{R^2}{u^2}\left (du^2 +u^2 d\Omega_5^2\right )\right ],
\end{equation}
and the dilaton and $B$ field
\begin{equation}
e^{2\phi}=\hat{g}_s^2, \ \ \ B_{-1}=-\alpha'\frac{u^4}{R^4}.
\end{equation}
This solution is the same as (\ref{lightsym}) up to a constant $b$.
The constant $b$ is not important since it can be produced by
rescaling $x_{\pm}$ as was pointed out in \cite{AOR}.

For other dimensions, the supergravity dual of NCOS is \cite{Harm1}
\begin{eqnarray}
&& ds^2= \alpha'\left [ H^{1/2}\frac{u^{7-p}}{R^{7-p}}
 \left (-dx_0^2 +dx_1^2\right) +H^{-1/2} (dx_2^2 +\cdots +dx_p^2 )
 + H^{1/2}(du^2 +u^2d\Omega_{8-p}^2 )\right], \nonumber\\
\label{ncosdual}
&& e^{2\phi} =\hat{g}_s^2 H^{(3-p)/2}\left(1+\frac{u^{7-p}}{R^{7-p}}\right),
 \ \ \ B_{01}=\alpha' \frac{u^{7-p}}{R^{7-p}}, \ \ \
 H = 1+\frac{R^{7-p}}{u^{7-p}},
\end{eqnarray}
Performing a Lorentz transformation on coordinates $x_0$ and $x_p$,
taking the infinite boost limit $\gamma \to \infty$ and rescaling as
(\ref{u}), we finally obtain
\begin{eqnarray}
&& ds^2=\alpha'\left [\left(\frac{u}{R}\right)^{(7-p)/2}\left (
  dx_+dx_- +dx_1^2 +\cdots +dx_{p-1}^2
 -\frac{ u^{7-p}}{R^{7-p}}dx_-^2 \right)
\right. \nonumber \\
&&~~~~~~~~\left. +\left(\frac{R}{u}\right)^{(7-p)/2}
    \left(du^2
 +u^2 d\Omega^2_{8-p}\right)\right], \nonumber\\
&& e^{2\phi} = \hat{g}_s^2 \left(\frac{R}{u}
 \right)^{(7-p)(3-p)/2}, \ \ \ B_{-1}=-\alpha' \frac{u^{7-p}}{R^{7-p}}.
\end{eqnarray}
Thus we arrive at again the supergravity dual of light-like
NCSYM, combining the supergravity dual of NCOS
and the Lorentz transformation.


\sect{NCSYM, OSYM and NCOS in the infinite-momentum frame}

In order to get the supergravity dual of light-like NCSYM, we have used
the Lorentz transformation along the nonisotropic
direction coordinates, such as $x_0$ and $x_3$, or $x_0$ and $x_p$.
In this section we consider the Lorentz transformation along the
isotropic directions. For generality, we start with the supergravity
dual (\ref{ncsymdual}) of spatial NCSYM in $p+1$ dimensions.
Obviously, the solution (\ref{ncsymdual}) is Lorentz invariant if one
performs a Lorentz coordinate transformation along one of directions
$x_1$ to $x_{p-2}$. To produce a nontrivial effect, we make a Lorentz
transformation along one of those directions, say $x_1$, for the
supergravity dual of finite temperature NCSYM. The latter is \cite{CO}
\begin{eqnarray}
&& ds^2=\alpha'\left [\left(\frac{u}{R}\right)^{(7-p)/2}\left (-
  f dx_0^2 +dx_1^2 +\cdots +dx_{p-2}^2
  +\tilde{h}(dx_{p-1}^2 +dx_p^2)\right)
   \right. \nonumber \\
&&~~~~~~~~\left. +\left(\frac{R}{u}\right)^{(7-p)/2}
    \left( f^{-1}du^2
 +u^2 d\Omega^2_{8-p}\right)\right], \nonumber\\
&& e^{2\phi} = \hat{g}_s^2 \tilde{h}\left(\frac{R}{u}
 \right)^{(7-p)(3-p)/2}, \ \ \ B_{p-1,p}
 = \frac{\alpha'}{\tilde{b}}\frac{(au)^{7-p}}{1+(au)^{7-p}},
\end{eqnarray}
where
\begin{eqnarray}
&& \tilde{h} = \frac{1} {1+(au)^{7-p}}, \ \ \ a^{7-p} = \tilde{b}^2/R^{7-p},
 \nonumber \\
&& f=1-\frac{u_0^{7-p}}{u^{7-p}}, \ \ \
R^{7-p} = \frac{1}{2}(2\pi)^{6-p}\pi^{-(7-p)/2}\Gamma[(7-p)/2] \hat{g}_s N.
\end{eqnarray}
We now make a Lorentz transformation for coordinates $x_0$ and $x_1$
and rescale the non-extremal parameter $u_0$ as
\begin{equation}
\label{P}
u_0^{7-p}e^{2\gamma} = P,
\end{equation}
where $P$ is a finite constant. When $\gamma \to \infty$, we have
\begin{eqnarray}
&& ds^2=\alpha'\left [\left(\frac{u}{R}\right)^{(7-p)/2}\left (
   dx_+dx_- +dx_2^2 +\cdots +dx_{p-2}^2
  + \tilde{h}(dx_{p-1}^2 +dx_p^2 )
   +\frac{P}{u^{7-p}}dx_-^2 )\right)
   \right. \nonumber \\
\label{pncsym}
&&~~~~~~~~\left. +\left(\frac{R}{u}\right)^{(7-p)/2}
    \left( du^2
 +u^2 d\Omega^2_{8-p}\right)\right],
\end{eqnarray}
and the dilaton and $B$ fields are kept unchanged, where $x_{\pm}=
x_1 \pm x_0$. This is a supergravity solution with a $pp$-wave,
and $P$ has an interpretation as the momentum density of the wave.
Let us notice that when the $B$ field vanishes, that is $\tilde{h}=1$,
the solution (\ref{pncsym}) reduces to
\begin{eqnarray}
&& ds^2=\alpha'\left [\left(\frac{u}{R}\right)^{(7-p)/2}\left (
   dx_+dx_- +dx_2^2 +\cdots +dx_{p-2}^2
  + dx_{p-1}^2 +dx_p^2
   +\frac{P}{u^{7-p}}dx_-^2 )\right)
   \right. \nonumber \\
\label{psym}
&&~~~~~~~~\left. +\left(\frac{R}{u}\right)^{(7-p)/2}
    \left( du^2
 +u^2 d\Omega^2_{8-p}\right)\right],
\end{eqnarray}
and the dilaton and NS $B$ field
\begin{equation}
e^{2\phi}= \hat{g}_s^2 \left (\frac{R}{u}\right)^{(7-p)(3-p)/2},
 \ \ \ B=0.
\end{equation}
Furthermore, when $p=3$ the above solution (\ref{psym}) goes to the one
obtained in \cite{Cveti}, there the authors of \cite{Cveti} have studied
the decoupling limit and near-horizon geometry of D3-branes (M2- and
M5-branes) with a $pp$-wave. They got the solution (\ref{psym}) with $p=3$,
(in this case the metric is of the Kaigorodov-type and preserves 1/4
supersymmetries), and argued that this is the supergravity dual of the
ordinary (3+1)-dimensional SYM in an infinitely-boosted frame with constant
momentum density. In this sense, the solution (\ref{psym}) gives a
generalization of \cite{Cveti}, which describes the supergravity dual of the
ordinary $(p+1)$-dimensional SYM in an infinitely-boosted frame. It is easy
to show that the solution (\ref{psym}) can also be obtained by taking the
decoupling limit for the (D$p$, W) bound states, where W denotes a $pp$-wave.

Thus it is natural to regard the solution (\ref{pncsym}) as the supergravity
dual of spatial NCSYM in an infinitely-boosted frame and such a
solution (\ref{pncsym}) can be achieved through considering the NCSYM limit
for the (D$(p-2)$, W, D$p$) bound states \cite{Lu1,OZ}.

Next let us consider the finite temperature picture for the supergravity
dual of NCOS in $p+1$ dimensions \cite{Harm1}
\begin{eqnarray}
&& ds^2= \alpha'\left [ H^{1/2}\frac{u^{7-p}}{R^{7-p}}
 \left (-f dx_0^2 +dx_1^2\right) +H^{-1/2} (dx_2^2 +\cdots +dx_p^2 )
 + H^{1/2}(f^{-1}du^2 +u^2d\Omega_{8-p}^2 )\right], \nonumber\\
&& e^{2\phi} =\hat{g}_s^2 H^{(3-p)/2}\left(1+\frac{u^{7-p}}{R^{7-p}}\right),
 \ \ \ B_{01}=\alpha' \frac{u^{7-p}}{R^{7-p}}, \ \ \
 H = 1+\frac{R^{7-p}}{u^{7-p}}.
\end{eqnarray}
Again, making a Lorentz transformation for coordinates $x_0$ and $x_1$,
and rescaling as (\ref{P}), we have
\begin{eqnarray}
 ds^2 &=& \alpha'\left [ H^{1/2}\frac{u^{7-p}}{R^{7-p}}
 \left ( dx_+dx_- + \frac{P}{u^{7-p}}dx_-^2\right)
 +H^{-1/2} (dx_2^2 +\cdots +dx_p^2 )
 \right. \nonumber \\
 & &~~~~~~ \left. + H^{1/2}(du^2 +u^2d\Omega_{8-p}^2 )\right],
\end{eqnarray}
and the dilaton and NS $B$ field unchanged.
This supergravity solution has a good
interpretation as the gravity dual of NCOS in an infinitely-boosted frame.
Of course, this solution can also be reached by considering the NCOS limit
for (F1, W, D$p$) bound states \cite{Lu2}.

Thus we have given the supergravity duals for the OSYM, spatial NCSYM, and
the NCOS in an infinitely-boosted frame by combining the SYM, NCSYM, and
NCOS limits, and the Lorentz transformation, respectively. We note that these
supergravity duals look like the one for light-like NCSYM in the sense that
they all have the $pp$-wave component in solutions, but actually they are
quite different.

\sect{Conclusions}

Combining the Lorentz transformation and the spatial NCSYM limit, or the
space-time NCOS limit, we have ``derived'' the light-like NCSYM limit.
The light-like NCSYM is unitary. We have argued that the SYM with space-time
noncommutativity, which comes from the Lorentz transformation of the
spatial NCSYM, is unitary as well. Actually, the two theories
connected through the Lorentz transformation are equivalent. Perhaps the
spatial NCSYM gives a better description. Along the consideration
of light-like NCSYM limit, we obtain the supergravity dual
of the light-like NCSYM using the Lorentz transformation and the gravity
dual of spatial NCSYM or of the NCOS, both giving the same result.
These supergravity duals are the same as those given by
AOR \cite{AOR}, where the authors first constructed the D$p$-brane solution
with light-like NS $B$ field and then took a usual SYM decoupling limit.

In addition, as a comparison, we have given supergravity duals for
the OSYM, spatial NCSYM, and the NCOS in an
infinitely-boosted frame with finite momentum density, by using the
Lorentz transformation and a certain rescaling. These supergravity
duals can also be obtained through considering certain decoupling limits
for the (D$p$, W), (D$(p-2)$, W, D$p$) and (F1, W, D$p$) bound states,
respectively.

\section*{Acknowledgments}

This work was supported in part by the Japan Society for the Promotion of
Science and in part by Grants-in-Aid for Scientific Research
Nos. 99020, 12640270 and in part by Grant-in-Aid on the Priority Area:
Supersymmetry and Unified Theory of Elementary particles.

\newcommand{\ATMP}[1]{Adv.\ Theor.\ Math.\ Phys.\ {\bf #1}}
\newcommand{\NP}[1]{Nucl.\ Phys.\ {\bf #1}}
\newcommand{\PL}[1]{Phys.\ Lett.\ {\bf #1}}
\newcommand{\PR}[1]{Phys.\ Rev.\ {\bf #1}}
\newcommand{\IJMP}[1]{Int.\ Jour.\ Mod.\ Phys.\ {\bf #1}}
\newcommand{\JHEP}[1]{J.\ High\ Energy\ Phys.\ {\bf #1}}

\end{document}